\documentclass[prc,twocolumn,nofootinbib,superscriptaddress]{revtex4-1}
\usepackage{graphicx,amsmath,amssymb,bm,tabularx}
\usepackage{multirow,dcolumn}

\usepackage{xcolor}

\newcolumntype{.}{D{.}{.}{2.3}}

\begin{document}

\title{$^{10}$Be-nucleus optical potentials developed from chiral effective field theory $NN$ interactions}

\author{V.\ Durant}
\email[Email:~]{vdurant@uni-mainz.de}
\affiliation{Institut f\"ur Kernphysik, Johannes Gutenberg-Universit\"at Mainz, D-55099 Mainz, Germany}

\author{P.\ Capel}
\email[Email:~]{pcapel@uni-mainz.de}
\affiliation{Institut f\"ur Kernphysik, Johannes Gutenberg-Universit\"at Mainz, D-55099 Mainz, Germany}
\affiliation{Physique Nucl\'eaire et Physique Quantique (CP 229),Universit\'e libre de Bruxelles (ULB), B-1050 Brussels, Belgium}

\begin{abstract}
We present a determination of optical potentials for $^{10}$Be-nucleus collisions using the double-folding method to compute the real part and Kramers-Kronig dispersion relations to derive the imaginary part. 
As microscopic inputs we use chiral effective field theory nucleon-nucleon interactions at next-to-next-to-leading order combined with state-of-the-art nucleonic densities. With these potentials, we compute elastic scattering cross sections for the exotic nucleus $^{10}$Be off various targets, and compare them to experiment. Without any fitting parameter, we obtain good agreement with data. For collisions on light targets, we observe significant uncertainty related to the short-range physics, whereas for heavy targets that uncertainty remains small. 
\end{abstract}

\maketitle
\section{Introduction}
One of the most important inputs in the study of nuclear reactions is the interaction between the colliding nuclei~\cite{Bran97heavyion}. Typically, the nuclear part of these interactions is described by phenomenological optical potentials (POPs), whose parameters are fitted to experimental data. Given their nature, they lack predictive power and cannot be used to describe reactions for which no data are available. To reduce uncertainty, it would be useful to derive optical potentials from first principles. This becomes more and more crucial with the development of new radioactive-ion beam facilities, where the structure of exotic nuclei are mostly studied through nuclear reactions~\cite{Baye12HaloNuclei}.

In the case of nucleon-nucleus potentials, there have been several approaches using $NN$ forces from chiral effective field theory (EFT)~\cite{Vora18OpPot,Idin19OpPot,Rotu20OpPot,White20OpPot}. These $NN$ interactions are expressed as an order-by-order expansion, which allows for a systematic improvement of the description of observables~\cite{Mach11PR,Hamm13RMP}. All these studies give interesting results, but are focused on the collision of a nucleon with a target nucleus. To describe nucleus-nucleus interactions, there have been efforts using different kinds of microscopic interactions within the double-folding formalism~\cite{Satc79Folding}. In this formalism, the potentials between nuclei are determined from two fundamental inputs: realistic nuclear densities and microscopic $NN$ interactions. In this way, either the real part of the optical potential~\cite{Cham02SPdens,Pere09ImDFPot} or both its real and imaginary parts~\cite{Furu12OpPot,Mino15ChEFTCC,Khoa16DFM} can be determined.
  
In our previous studies~\cite{Dura17DFP,Dura20ImDisp,Dura20KKrel}, we have combined these two ideas. We have taken $NN$ interactions developed within a chiral EFT framework to construct nucleus-nucleus optical potentials using the double-folding method.  To construct the imaginary part, we have suggested to use the dispersion relations~\cite{Dura20ImDisp,Dura20KKrel}. This enabled us to obtain optical potentials without any fitting parameter. This approach allowed us to satisfactorily describe elastic scattering and low-energy fusion involving light stable projectiles such as $\alpha$, $^{12}$C, and $^{16}$O with a variety of targets ranging from $\alpha$ to $^{120}$Sn.

Our goal for the present study is to extend this method away from stability. We concentrate on reactions involving $^{10}$Be, which is radioactive and exhibits deformation and cluster-like configurations~\cite{AlKha06Halo10Be}.
The validity of our method would enable us to construct reliable nucleus-nucleus interactions that are key inputs of few-body models of reactions~\cite{Baye12HaloNuclei}. In particular, $^{10}$Be-target interactions could then be used in the description of reactions with $^{11}$Be, the archetypical one-neutron halo nucleus. Studies involving halo nuclei are an active topic of research for both nuclear-reaction and structure communities. Having the ability to construct optical potentials from first principles would be an asset to these studies.

This paper is organized as follows: in Sec.~\ref{sec:optical} we give a brief overview of the formalism of the double-folding technique and the ways of building the imaginary part of the optical potential. In Secs.~\ref{sec:12C},~\ref{sec:208Pb}, and~\ref{sec:64Zn} we present results for elastic scattering of $^{10}$Be on different targets at energies at which experimental data exists: $^{12}$C at $E_\text{lab}=595$ MeV~\cite{Cort97Be10C12}, $^{208}$Pb at $E_\text{lab}=127$ MeV~\cite{Duan20Be10Pb208}, and $^{64}$Zn at $E_\text{lab}=28.3$ MeV~\cite{DiPie12Be1064Zn}. Finally, we summarize and give an outlook in Sec.~\ref{sec:sum}.

\section{Optical potentials from the double-folding formalism}
\label{sec:optical}
To analyse elastic scattering within the optical model, the nuclear part of the interaction between the colliding nuclei is described by a complex potential.   
For the real part, we assume the antisymmetrized double-folding potential (DFP), V$_\text{F}$, constructed as the sum of a direct (D) and an exchange (Ex) contributions: $V_\text{F}=V_\text{D}+V_\text{Ex}$. For a detailed explanation, see Refs.~\cite{Furu12OpPot,Dura17DFP}. The direct part is the average of the $NN$ interaction $v$ over the nucleonic densities
\begin{equation}
V_\text{D}(r) = \sum_{i,j = n,p} \iint \rho^i_1({\bf r}_1) \, v^{ij}({\bf s}) \, \rho^j_2({\bf r}_2)
\, d^3{\bf r}_1 d^3{\bf r}_2 \,,
\label{eq:direct}
\end{equation}
\noindent where ${\bf r}$ is the relative coordinate between the centers of mass of the nuclei, ${\bf r}_{1}$ and ${\bf r}_{2}$ are the inner coordinates of nucleus 1 and 2, respectively; ${\bf s}={\bf r}-{\bf r}_1+{\bf r}_2$ is the relative coordinate between any given pair of points in the projectile and target, and $\rho_{j}^{i}$ are the neutron ($i=n$) and proton ($i=p$) density distributions of nucleus $j$.
The exchange part of the potential arises from the fact that identical nucleons in the projectile and the target cannot be distinguished from one another. It reads
\begin{multline}
V_\text{Ex}(r,E_\text{c.m.}) = \sum_{i,j = n,p} \iint \rho^i_1({\bf r}_1,{\bf r}_1+{\bf s})
\, v^{ij}_\text{Ex}({\bf s}) \\
\times \rho^j_2({\bf r}_2,{\bf r}_2-{\bf s}) \exp \left[\frac{i{\bf k}(r)\cdot{\bf s}}{\mu/m_N}\right] \,
d^3{\bf r}_1 d^3{\bf r}_2 \,, \label{eq:exchange}
\end{multline}
\noindent where $\mu$ is the reduced mass of the colliding system, $v_\text{Ex}=-P_{12}v$ is the exchange contribution from the $NN$ potential, the integral runs over the density matrices $\rho^i_{1,2}({\bf r},{\bf r} \pm {\bf s})$ of the nuclei, and the momentum for the nucleus-nucleus relative motion ${\bf k}$ is given by
\begin{equation}
k^2(r)=\frac{2\mu}{\hbar^2} \, \Bigl[ E_\text{c.m.} - V_\text{F}({r},E_\text{c.m.}) - V_\text{Coul}({r}) 
\Bigr] \,. \label{eq:k}
\end{equation}
Due to the dependence of ${\bf k}$ on the double-folding potential $V_\text{F}$, $V_\text{Ex}$ has to be determined self-consistently.

Following Refs.~\cite{Dura17DFP,Dura20ImDisp,Dura20KKrel}, we use as $NN$ potential local chiral EFT intreractions. These are based on those of Refs.~\cite{Geze13QMCchi,Geze14long}, which are derived and regulated directly in $r$-space. To test the sensitivity of our calculation to short-range physics, we apply different regulators, with cutoffs $R_0=1.2$ and 1.6~fm~\cite{Dura17DFP}.

To describe the absorptive imaginary part of the potential, we have explored two possibilities: the first one is a zeroth-order approximation setting the imaginary part proportional to the real DFP, as suggested in Refs.~\cite{Alv03ImPot,Pere09ImDFPot}
\begin{equation}
W=N_WV_F\,.
\label{eq:NW}
\end{equation}

The second possibility is using Kramers-Kronig relations, better known in our field as \emph{dispersion relations}, which link the real and imaginary parts of the interaction~\cite{Carl89Dispersion,Gonz01DisRel}. These relations are the application of the Sokhotski-Plemelj theorem and they relate the imaginary part of the potential $W$ with the energy-dependent part of the DFP through~\cite{Dura20KKrel}
\begin{equation}
W(r,E_\text{c.m.})=-\frac{1}{\pi}\mathcal{P}\int_{-\infty}^{+\infty} \frac{V_\text{Ex}(r,E)}{E-E_\text{c.m.}}dE\,,
\label{eq:W_Disp}
\end{equation}
\noindent where $\mathcal{P}$ represents the principal value integral.
Contrary to Eq.~(\ref{eq:NW}), this approach provides an efficient constraint on the imaginary term of the nucleus-nucleus interactions without involving any free parameter. However, it does not include couplings to single excited states of either of the nuclei. At sufficiently high energy (higher than the Coulomb barrier) this approach is justified and leads to good agreement with data for both closed and non-closed shell nuclei~\cite{Dura20ImDisp,Dura20KKrel}.

The double-folding potentials constructed in this way exhibit a systematic order-by-order behavior expected in EFT, very similar to the one seen in Ref.~\cite{Dura17DFP}. These features carry through to the elastic scattering cross sections. For this reason, in the following we will show only results at N$^2$LO, which reproduce better the experimental data. We limit our calculations to this order, because we concentrate on the description of elastic-scattering, for which the inclusion of orders beyond N$^2$LO does not have much impact~\cite{Vor17OpPotN4LO}.

\section{$^{10}$Be-$^{12}$C scattering}
\label{sec:12C}
\begin{figure}[b]
\begin{center}
\includegraphics[width=0.95\columnwidth]{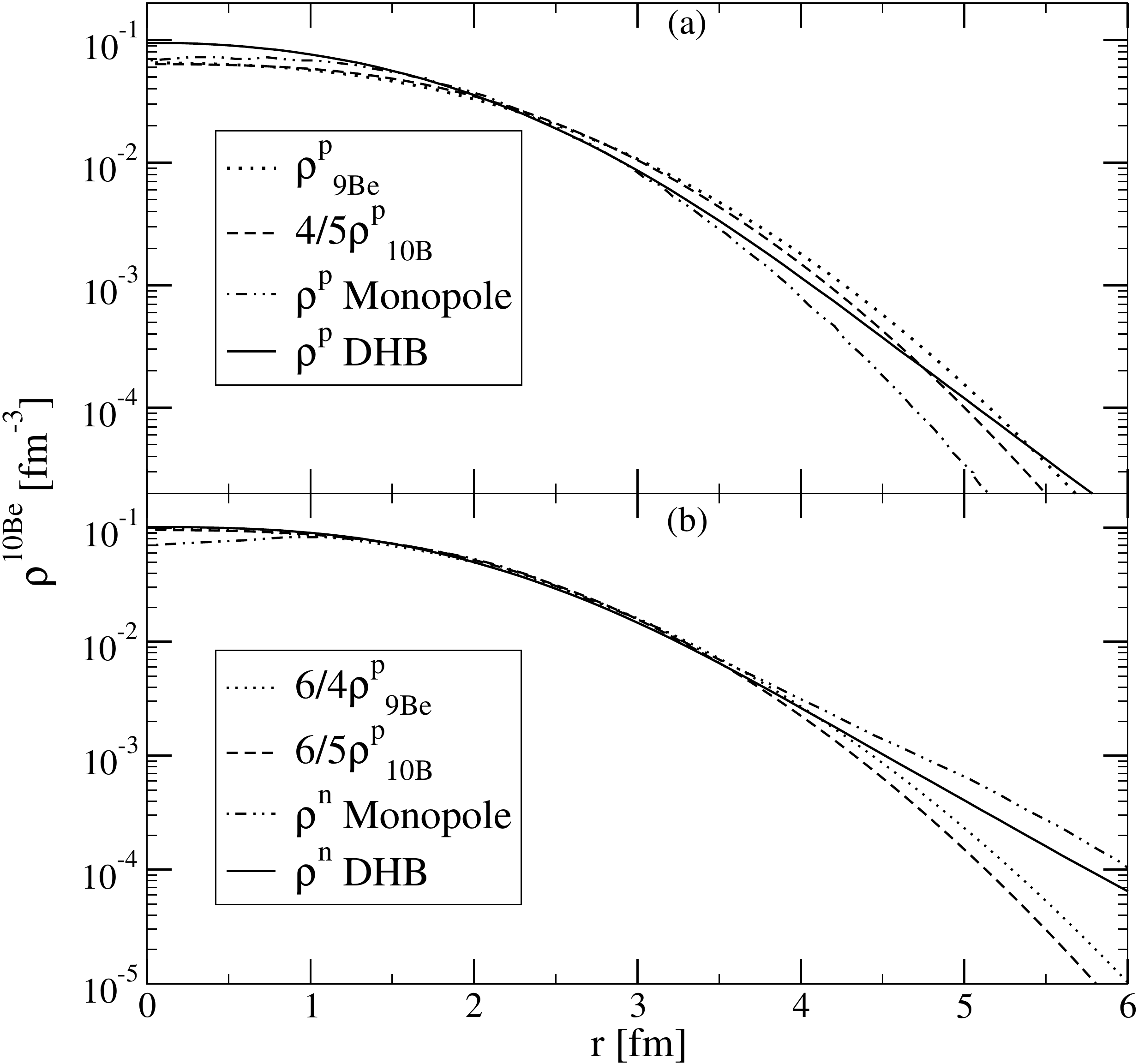}
\caption{$^{10}$Be density profiles for (a) protons, and (b) neutrons. Four types of densities are shown: scaled harmonic oscillator densities~\cite{Devr87rhoel} for $^{9}$Be (dotted lines) and  $^{10}$B (dashed lines), monopole densities~\cite{Desc19BeCluster} (dashed-dotted lines), and the densities calculated using DHB theory~\cite{Cham21REGINA} (solid lines).} 
\label{fig:10Be_rho}
\end{center}
\end{figure}

We start this study with the analysis of $^{10}$Be-$^{12}$C elastic scattering at $E_\text{lab}=$ 595 MeV. We consider the experimental data as well as the POP fitted to them from Ref.~\cite{Cort97Be10C12}. 
First, we assess the impact the nuclear densities on the elastic-scattering cross sections. Figure~\ref{fig:10Be_rho} shows the  densities for the $^{10}$Be (a) proton and (b) neutron distributions. We use two different kinds of profiles: either fitted from experimental data, or obtained from microscopic calculations. In Ref.~\cite{Devr87rhoel}, charge densities parametrised as harmonic oscillator functions (HO) are fitted from electron scattering off $^9$Be and $^{10}$B. From them we infere proton densities, which we then scale to the number of protons and neutrons of $^{10}$Be. These scaled distributions are plotted as dotted and dashed lines in Fig.~\ref{fig:10Be_rho}. We also explore two microscopic descriptions: monopole densities obtained from cluster calculations~\cite{Desc19BeCluster}, and Dirac-Hartree-Bogoliubov (DHB) results given by the code REGINA~\cite{Cham21REGINA} (dashed-double-dotted and solid lines in Fig.~\ref{fig:10Be_rho}). 

\begin{figure}[]
\begin{center}
\includegraphics[width=0.95\columnwidth]{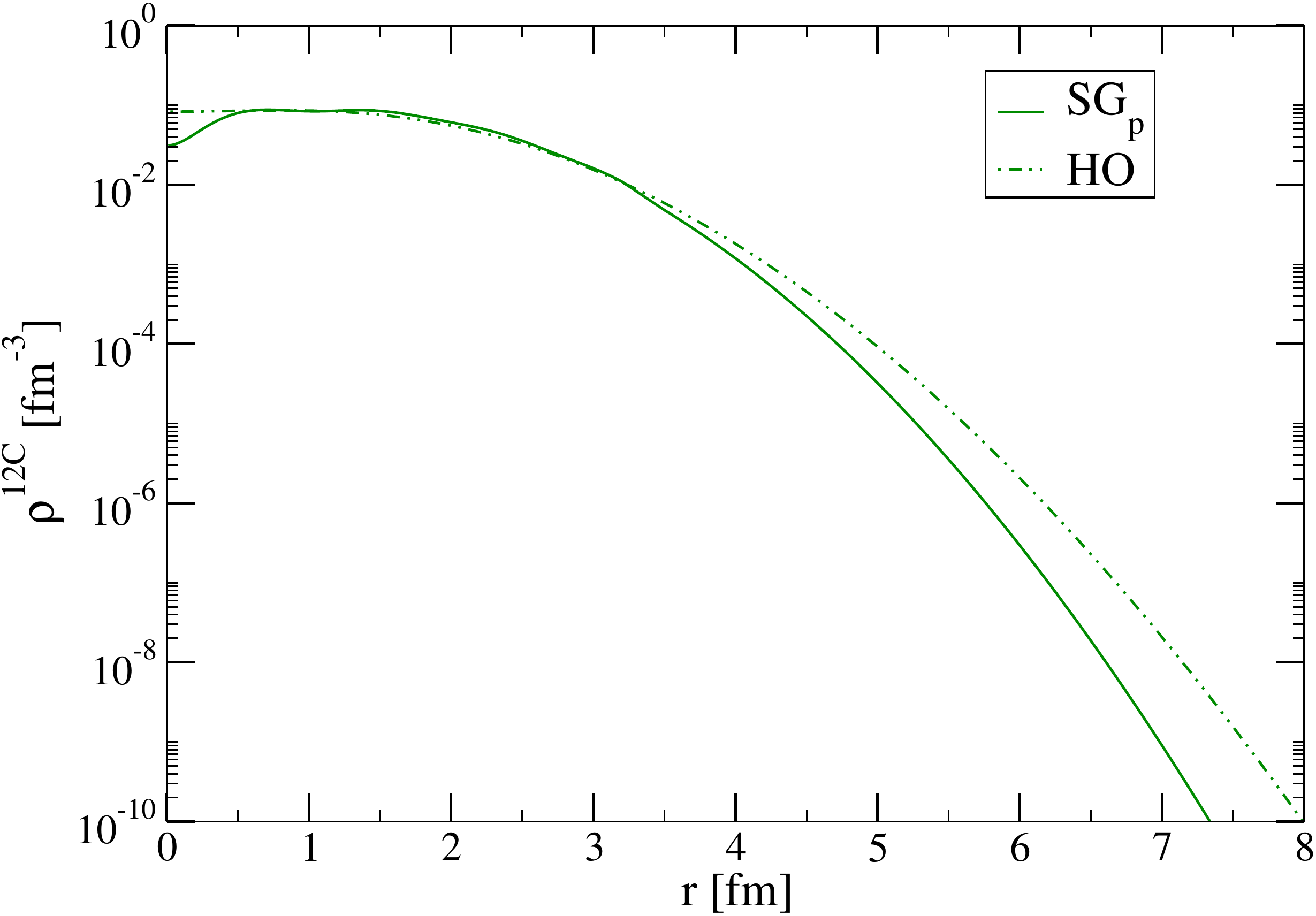}
\caption{Proton density profiles for $^{12}$C: the Sum-of-Gaussians proton particle density (SG$_\text{p}$, solid line), and the harmonic oscillator parametrization (HO, dashed-double-dotted line)~\cite{Devr87rhoel}.} 
\label{fig:12C_rho}
\end{center}
\end{figure}

For $^{12}$C, we make the approximation $\rho^{p}=\rho^{n}$, since it is a light and stable nucleus with equal number of protons and neutrons. Two different proton densities can be seen in Fig.~\ref{fig:12C_rho}. First, a Sum-of-Gaussians nucleonic density (SG$_\text{p}$; solid lines), based on the parametrisation of the charge density obtained through electron scattering in Ref.~\cite{Devr87rhoel}. Second, a density profile obtained through electron scattering parametrised as a harmonic oscillator function (dashed-double-dotted lines)~\cite{Devr87rhoel}.

\begin{figure}[]
\begin{center}
\includegraphics[width=0.95\columnwidth]{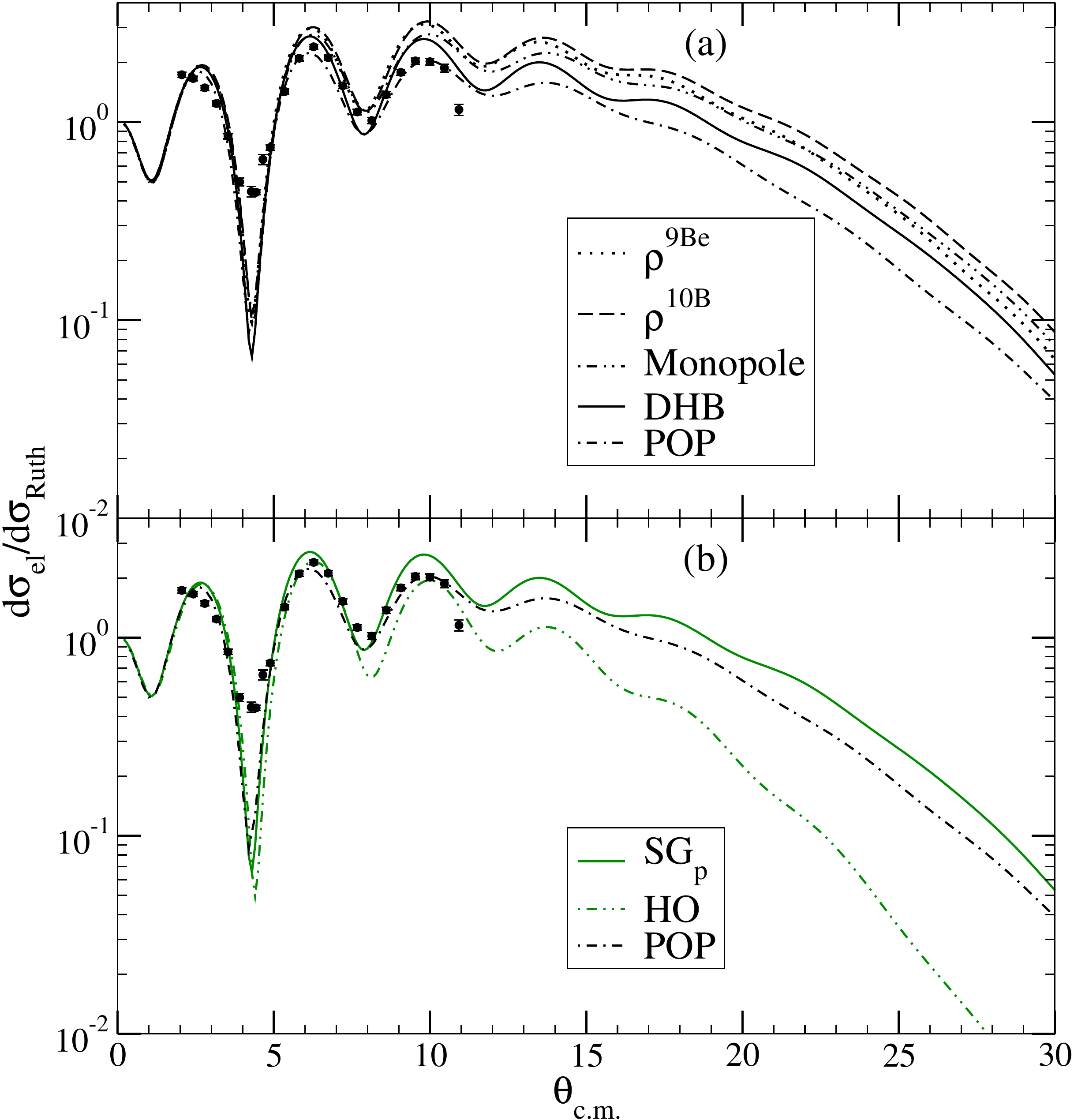}
\caption{$^{10}$Be-$^{12}$C elastic-scattering cross sections (normalized to Rutherford) at $E_\text{lab}=595$ MeV as a function of the center-of-mass angle. Impact of (a) the $^{10}$Be and (b) $^{12}$C densities.  For panel (a) and (b) the line types correspond to those described in the caption of Figs.~\ref{fig:10Be_rho} and~\ref{fig:12C_rho}, respectively. In all cases, the $NN$ cutoff is $R_0=1.6$ fm and the imaginary part was calculated through Kramers-Kronig relations. Experimental data from Ref.~\cite{Cort97Be10C12}.} 
\label{fig:10Be12C_CS_rho_DRel}
\end{center}
\end{figure}

Figure~\ref{fig:10Be12C_CS_rho_DRel} shows the cross sections for elastic scattering normalized to Rutherford using the different densities for the collision of $^{10}$Be and $^{12}$C at $E_\text{lab}=595$ MeV. The black dashed-dotted line shows the reference cross section obtained with the POP. All our results are calculated with the $NN$ cutoff $R_0=1.6$ fm and the Kramers-Kronig relations to determine the imaginary part of the optical potentials [Eq.~(\ref{eq:W_Disp})]. The influence of $R_0$ and the imaginary part is studied and discussed later in this section. Panel (a) shows the dependence of the elastic-scattering cross sections on the $^{10}$Be density. For the $^{12}$C density, we use SG$_\text{p}$, which has been shown to provide the best results for scattering involving $^4$He~\cite{Dura20KKrel}. From these results, we see that up to $\approx 7^{\circ}$ all the densities give cross sections that are in good agreement with the data from Ref.~\cite{Cort97Be10C12}. At larger angles, these distributions lead to cross sections with different magnitudes, but remain in phase with the oscillations of the experimental data. Between the second and third maxima, there is not much spread amongst the cross sections obtained with different densities. At larger angles, DHB calculations (solid line) agree slightly better with the POP results, so we will use this profile to assess the impact of the $^{12}$C density in the cross sections. 

In panel (b) we can see the results obtained with the different $^{12}$C distributions: SG$_\text{p}$ gives cross sections that are in phase with experimental data, while results with the HO density are shifted towards larger angles, starting at the second minimum. We thus conclude that SG$_\text{p}$ is optimal in this case as well.
Having determined the best densities to reproduce the experimental data, we can assess the impact of the $NN$ interaction and the imaginary part of the potential. 

\begin{figure}[]
\begin{center}
\includegraphics[width=0.95\columnwidth]{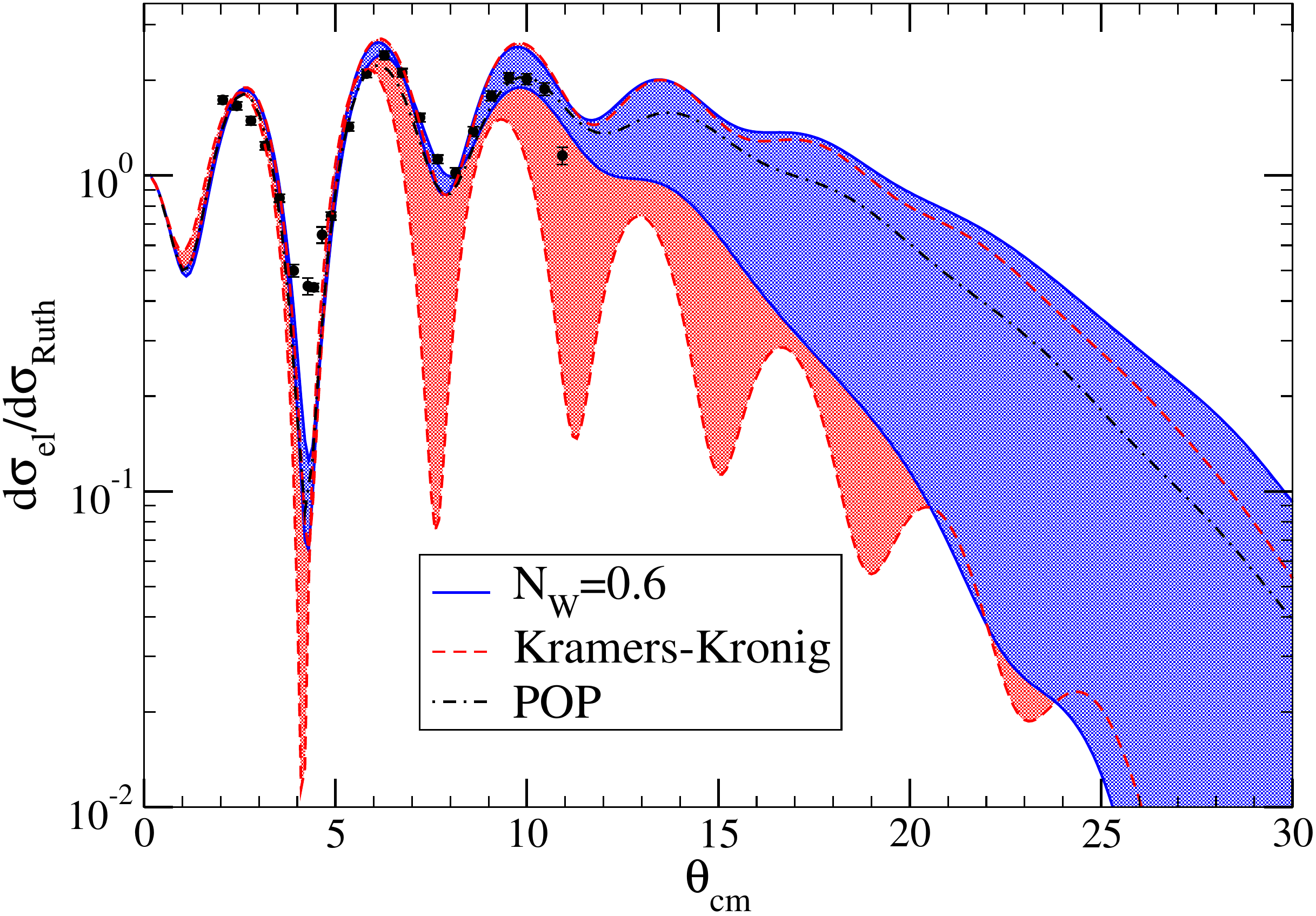}
\caption{$^{10}$Be-$^{12}$C elastic-scattering cross sections (normalized to Rutherford) at $E_\text{lab}=595$ MeV as a function of the center-of-mass angle. The bands show the $R_0=1.2$ -- 1.6 fm dependence. The red and blue bands show results with Kramers-Kronig relations [Eq.~(\ref{eq:W_Disp})] and $N_W=0.6$ [Eq.~(\ref{eq:NW})], respectively. Experimental data from Ref.~\cite{Cort97Be10C12}.} 
\label{fig:10Be12C_CS_DRel}
\end{center}
\end{figure}

In Fig.~\ref{fig:10Be12C_CS_DRel}, the bands depict the $R_0$ dependence, where the upper and lower lines correspond to the result using $R_0=1.6$ and 1.2 fm, respectively. The red band shows the results using Kramers-Kronig relations, while the blue band corresponds to calculations setting the imaginary part proportional to the real part with $N_W=0.6$ [Eq.~(\ref{eq:NW})]. For both imaginary parts, the $R_0$ dependence is small at forward angles, where the Coulomb interaction plays a major role. At larger angles, they have a sizable dependence on the $R_0$ cutoff, which indicates that short-range $NN$ physics becomes more relevant. This impact of the nuclear part of the potential is to be expected in the collision of two light nuclei that have small radii, as we have also observed in previous studies~\cite{Dura20ImDisp,Dura20KKrel}. The results for $R_0=1.6$~fm (upper lines) are in good agreement with the data for both choices of the imaginary part. Note that for this cutoff, the discrepancy between them is smaller than their difference with experiment. This is not the case for $R_0=1.2$ (lower lines), where the two imaginary potentials lead to very different cross sections. In particular, when using the Kramers-Kronig relations the cross section with $R_0=1.2$ fm (lower red dashed line) is not in phase with the data beyond the second minimum. This $NN$ potential contains more short-range information than the one with the larger cutoff $R_0=1.6$ fm. Accordingly, the unphysical oscillations obtained with the Kramers-Kronig relations suggest that the $NN$ interaction with $R_0$=1.2 fm is too hard to be used in this approach.
Let us stress that we set $N_W=0.6$ because it best fits the data. Although this value agrees with the $N_W$ range used in Ref.~\cite{Alv03ImPot}, the Kramers-Kronig relations provide an imaginary part without any fitting parameter.

\section{$^{10}$Be-$^{208}$Pb scattering}
\label{sec:208Pb}
Next we study a reaction at lower energy: $^{10}$Be-$^{208}$Pb elastic scattering at $E_\text{lab}=$ 127 MeV, which corresponds to the experimental conditions of Ref.~\cite{Duan20Be10Pb208}. To describe the $^{208}$Pb density, we use the relativistic mean field (RMF) profiles from Ref.~\cite{Chen15NStars}. As we found also in our previous work~\cite{Dura20KKrel}, since $^{208}$Pb is a heavy nucleus, the sensitivity of our calculations is mostly dependant on the choice of the density of the light projectile $^{10}$Be. We have observed that for this collision the best $^{10}$Be density to reproduce the experimental data is also the profile given by DHB calculations, as it was in the $^{10}$Be-$^{12}$C case.

\begin{figure}[t]
\begin{center}
\includegraphics[width=0.95\columnwidth]{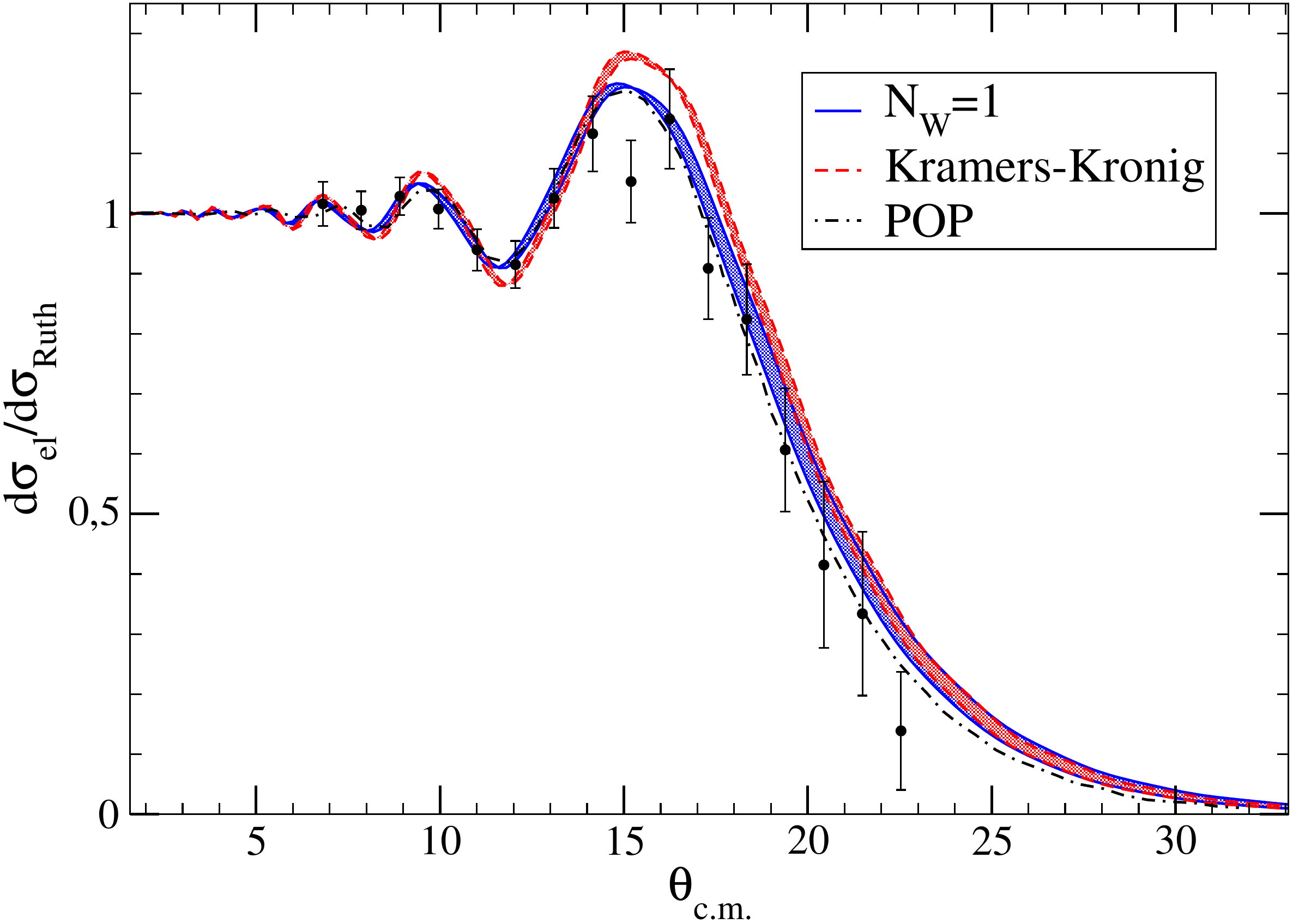}
\caption{$^{10}$Be-$^{208}$Pb elastic-scattering cross sections (normalized to Rutherford) at $E_\text{lab}=127$ MeV as a function of the center-of-mass angle. The bands show the $R_0=1.2$ -- 1.6 fm dependence. The red and blue bands show results with Kramers-Kronig relations [Eq.~(\ref{eq:W_Disp})] and $N_W=1.0$ [Eq.~(\ref{eq:NW})], respectively. Experimental data from Ref.~\cite{Duan20Be10Pb208}.} 
\label{fig:10Be208Pb_CS}
\end{center}
\end{figure}

Figure~\ref{fig:10Be208Pb_CS} compares our calculation with the data of Ref.~\cite{Duan20Be10Pb208}. It shows the dependence of the elastic-scattering cross sections on $R_0$ and the imaginary part of the potential. For the imaginary part we use Kramers-Kronig relations (red band), and $N_W=1$ (blue band), value that best reproduces the data. This choice is in agreement with the recommendation of Ref.~\cite{Alv03ImPot}, where they use higher values of $N_W$ for collisions that involve heavier nuclei. The black dashed-dotted line shows the results obtained with the POP given in Ref.~\cite{Duan20Be10Pb208}, which was fitted to the experimental data. 
The $R_0$ bands are quite narrow, especially compared to those in Fig.~\ref{fig:10Be12C_CS_DRel}. 
This lighter dependence on the $NN$ interaction is probably due to the dominance of the Coulomb interaction in the reaction, which is the same effect that we saw at small angles in $^{10}$Be-$^{12}$C scattering.
We want to remind the reader that the value of $N_W$ was chosen to best reproduce the data, and that is why in Fig.~\ref{fig:10Be208Pb_CS} the results using $W=V_\text{F}$ agree better with experiment than those obtained with Kramers-Kronig relations. Nevertheless, using the Kramers-Kronig relations, we still get good agreement with data and the reference optical potential without the need of any parameter, which is a confirmation that our approach gives a good description of the imaginary potential.

\section{$^{10}$Be-$^{64}$Zn scattering}
\label{sec:64Zn}

In this section we study the elastic-scattering cross sections for $^{10}$Be-$^{64}$Zn scattering at $E_\text{lab}=28.3$ MeV~\cite{DiPie12Be1064Zn}.
As in the previous section, cross sections for this collision are mostly dependent on the $^{10}$Be density, and not on the density of the heavier target. In this study we use once again the distributions from DHB calculations for $^{10}$Be. 
We use the results of the code REGINA~\cite{Cham21REGINA} for the $^{64}$Zn density.

\begin{figure}[]
\begin{center}
\includegraphics[width=0.95\columnwidth]{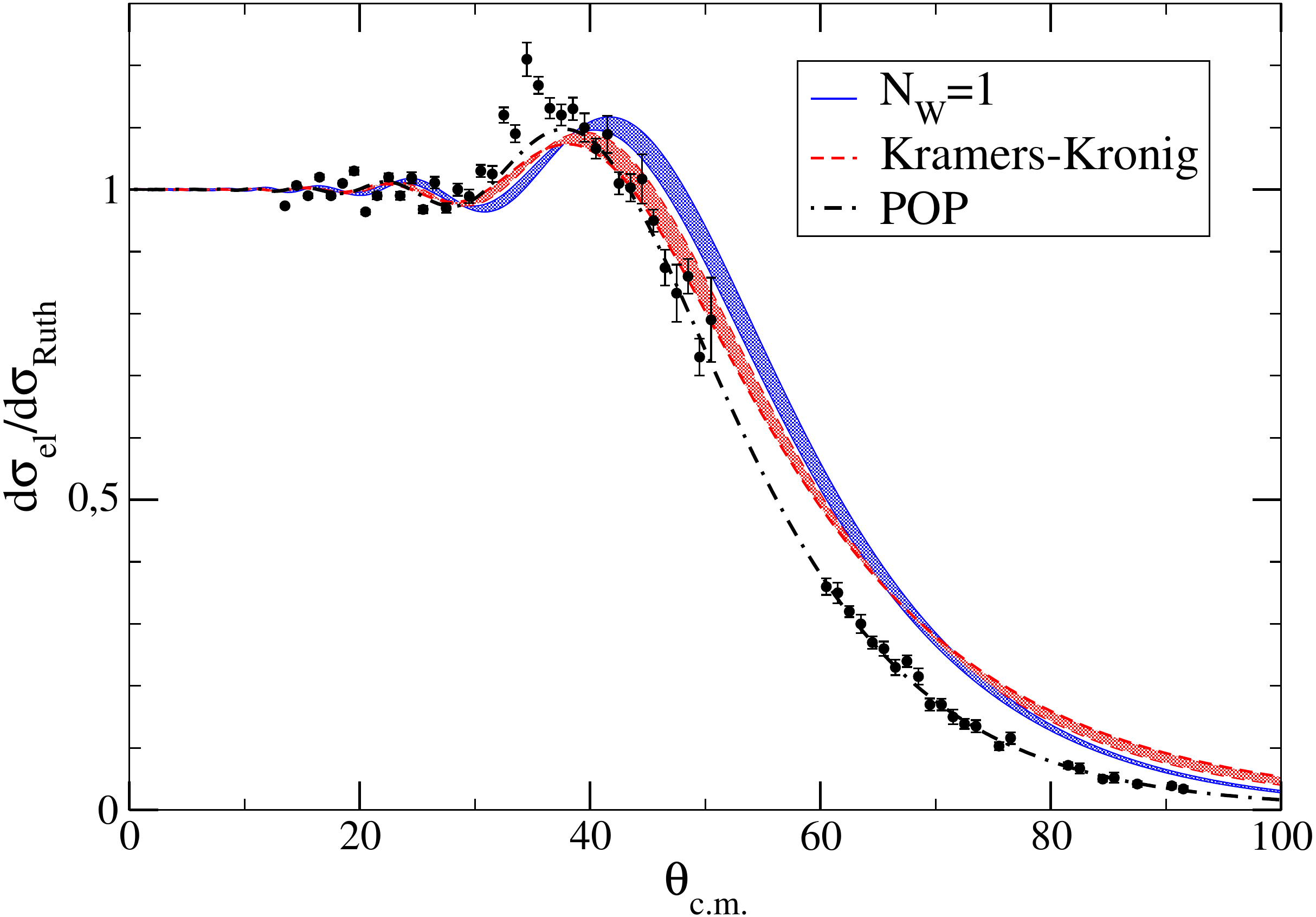}
\caption{$^{10}$Be-$^{64}$Zn elastic scattering cross sections (normalized to Rutherford) at $E_\text{lab}=28.3$ MeV as a function of the center-of-mass angle. The bands show the $R_0=1.2$ -- 1.6 fm dependence. The red and blue bands show results with Kramers-Kronig relations [Eq.~(\ref{eq:W_Disp})] and $N_W=1.0$ [Eq.~(\ref{eq:NW})], respectively. Experimental data from Ref.~\cite{DiPie12Be1064Zn}.} 
\label{fig:10Be64Zn_CS_DRel}
\end{center}
\end{figure}

In Fig.~\ref{fig:10Be64Zn_CS_DRel} we compare our results to the data of Ref.~\cite{DiPie12Be1064Zn}. We assess the $R_0$ and imaginary part dependences of the elastic-scattering cross sections. For the imaginary part we use Kramers-Kronig relations (red band), and $N_W=1$ (blue band). We can see that both descriptions give results that are shifted towards larger angles compared to the experiment and the POP of Ref.~\cite{DiPie12Be1064Zn} (dash-dotted line). In both cases, the $R_0$ band is quite narrow, since this is a low-energy Coulomb dominated reaction. The Kramers-Kronig relations agree better with data right after the maximum (at $\approx$ 40$^\circ$--50$^\circ$), but in general there is little difference with the results that we get using the $N_W$ prescription. As we discussed in Ref.~\cite{Dura20KKrel}, at low energies we expect a larger contribution from excitation to higher states, which will have an important effect in the imaginary part of the optical potential. Especially, the deformed nature of $^{64}$Zn and $^{10}$Be are likely to play an important role in the cross sections~\cite{DiPie19Be11Zn64,AlKha06Halo10Be}.

This system opens interesting new paths for our model. Our results suggest that at these lower energies excitations to low lying states have a significant impact and that beyond Hartree-Fock effects should be included. It will also be interesting to study the effect of the inclusion of deformation in the densities that we use to build our DFP. Those are questions that we plan to answer in the future.

\section{Conclusions and Outlook}
\label{sec:sum}
This work is an obvious extension of our previous studies~\cite{Dura17DFP,Dura20ImDisp,Dura20KKrel}. Its goal is to benchmark our approach in reactions involving more exotic projectiles. To this end, we have presented the derivation of $^{10}$Be-target interactions at different energies. To determine the real part, we used the double folding of local chiral EFT $NN$ interactions~\cite{Geze13QMCchi,Geze14long} over realistic nucleonic densities. We constrained the imaginary part of the optical potential using the Kramers-Kronig relations (also known as dispertion relations)~\cite{Carl89Dispersion,Gonz01DisRel}. 
Within this framework, we were able to reproduce data for the elastic scattering of $^{10}$Be off $^{12}$C, $^{208}$Pb, and $^{64}$Zn at different energies. For the collision off the light target $^{12}$C, we have seen that the densities of both nuclei play an important role in the results, as also does the $NN$ interaction. For heavier targets the main effect comes from the $^{10}$Be density. We have found that the dependence with the short-distance cutoff of the $NN$ interaction is large on the nuclear-dominated reaction $^{10}$Be-$^{12}$C, while it is small for the Coulomb-dominated ones. 

Using Kramers-Kronigs relations to constrain the imaginary part of the potential gives good results for high-energy collisions. These relations are not enough to fully describe the imaginary part of the optical potential at lower energies, where higher-order effects should be included in the model. This method provides reliable nucleus-nucleus optical potentials to describe reactions for which there are no data, in particular those that involve halo nuclei~\cite{Baye12HaloNuclei,Cap04Be11BU}. Knowing that there is no fitting or scaling parameter in our framework, these results clearly illustrate the interest of these potentials.

There remain several paths for improvement, at the level of both the many-body folding method and the input interactions. Accounting for the excited spectrum of the colliding nuclei would refine the description of the imaginary part of the optical potential through the application of dispersion relations to these energy-dependent terms. This would further improve our potentials and their description of the scattering processes, especially at low collision energies~\cite{Maha86DispRel}. Also, it would be interesting to study the impact of going beyond leading order in the density matrix expansion used in Eq.~(\ref{eq:exchange}), or the impact of using parametrizations of the diagonal densities that explicitely include deformation~\cite{Cham21REGINA}. 
Another aspect that needs to be investigated is the role of $3N$ interactions, as they also enter at N$^2$LO. In preliminary calculations for $^{16}$O-$^{16}$O~\cite{Webe17TFP}, we have observed that the contribution to the nucleus-nucleus potential arising from three-nucleon interaction is very small compared to the two-body contributions discussed here. Moreover, based on the results in nucleon-nucleus reactions~\cite{Vor20PotChEFT3N}, we expect these contributions to also be small in other systems. However, this needs to be further investigated. 
Finally, using \textit{ab initio} densities from chiral EFT would provide a consistent treatment of our inputs and would be a good step towards a unified description of structure and reactions.

\section*{Acknowledgments}
We thank J.\ Piekariewicz for the RMF density profiles. We also thank the International Atomic Energy Agency that provided the experimental data through their web page \href{www-nds.iaea.org}{www-nds.iaea.org}. 
This work was supported by the PRISMA$^+$ (Precision Physics, Fundamental Interactions and Structure of Matter) Cluster of Excellence, and 
Deutsche Forschungsgemeinschaft (DFG, German Research Foundation) and Projekt-ID 279384907 -- SFB 1245.

\bibliography{references}

\end{document}